\documentclass[reprint,prl,amsmath,amssymb,showpacs,twocolumn,footinbib]{revtex4-1}
\usepackage{amssymb,amsmath,amsfonts,placeins,psfrag,graphicx,epsfig,mathtools}
\usepackage{color}
\usepackage[hidelinks]{hyperref}

\newcommand{\rd}{\mathrm{d}}
\newcommand{\pd}[2]{\frac{\partial #1}{\partial #2}}

\newcommand{\eps}{\epsilon}

\newcommand{\beq}{\begin{equation}}
\newcommand{\eeq}{\end{equation}}

\DeclareMathOperator{\sinc}{sinc}

\begin{document}
\title{
Dynamical elastic contact of a rope with the ground} 
\author{Gregory Kozyreff}
\email{gkozyref@ulb.ac.be}
\author{Beno\^it Seron}
\affiliation{Optique Nonlin\'eaire Th\'eorique, Universit\'e libre de Bruxelles (U.L.B.), CP 231, Campus de la Plaine, 1050 Bruxelles, Belgium.}
\date{\today}

\begin{abstract}
A rope laid on the ground with one end subjected to time-dependent forcing is proposed as a prototypical elastic dynamical contact problem, which we study analytically, numerically, and experimentally. The dynamics is governed by an infinite set of linear and nonlinear resonances. In the limit of weak bending stiffness, the fundamental frequency is found to be independent of the rope tension. A transition between a radiation-less and a wave radiating state occurs via a series of grazing bifurcations, whereby new contacts between the rope and the ground are formed. The grazing bifurcations form overlapping Arnold tongues in the frequency-amplitude parameter space. Finally, for ropes with large bending stiffness and when the geometric nonlinearity is important, bistability is observed between several wave-making regimes.
\end{abstract}

\maketitle

Take a long piece of rope, lay it on the ground and give it a sudden vertical shake at one extremity. If that was vigorous enough, you are likely to see a bell-shape elevation travelling far away along the rope, until internal friction makes it disappear. Such a wave is familiar to anyone and simple-looking, but the underlying physics is more complicated than meets the eyes. In this Letter, we shed some light on the dynamics underlying wave generation by a combination of analytical, numerical, and experimental results. Aside from its recreational aspect, this problem is an opportunity to gather basic qualitative and quantitative knowledge into the theory of dynamical contacts involving deformable bodies. Indeed, questions of this kind can usually be approached only numerically~\cite{Wriggers-2006,Doyen-2011}.  Even in the static case, analytically tractable elastic contact problems are rare and most often proceed from Hertz's famous study of two spheres pressed against one another~\cite{Hertz-1896}. On the other hand,   the present problem has a practical interest, as it may be connected to some aspects of cable laying on ocean beds~\cite{Zajac-1957,Jawed-2014}, rails deformation under a moving load~\cite{Weitsman-1971}, the intrusion of a rod into a cylinder in the context of oil well drilling~\cite{Miller-2015b}, or parasitic contacts in rotating machines~\cite{Mokhtar-2017}. Importantly, and contrary to other systems combining vibration and impact~\cite{Astashev-2001,Murphy-2002}, as in the atomic force microscope in the taping regime~\cite{Dick-2008}, the contact point is not known in advance and becomes non-unique on the occasion of wave emission.

\begin{figure}
\includegraphics[width=8cm]{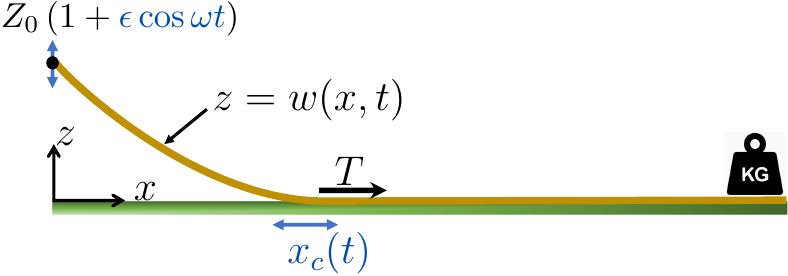}
\caption{Schematic of the rope with one end vibrated with relative amplitude $\eps$ and pulsation $\omega$ around a fixed elevation $Z_0$. Prior to vibrating the rope, a tension $T$ is applied and the resulting static deformation is maintained by a weight placed at the other end of the rope. The rope is sufficiently long to neglect reflection  from the weight. For $\eps$ sufficiently small, only one coordinate, $x_c(t)$, describes the edge of the contact region between the rope and the ground (contact point).}\label{fig:schematics}
\end{figure}

In order to study wave generation in a systematic way, we set up an experiment, schematically depicted in Fig.~\ref{fig:schematics}, where a long rope is excited harmonically with a small amplitude. Intuitively, one would expect that below a certain threshold frequency, the lifted part of the rope does no more than gently following the motion imparted at the extremity and that, above, waves are radiated along the rope. We determine such a frequency and find, unexpectedly, that it is independent of the rope tension when bending stiffness is negligible. Further analysis reveals that wave generation starts at grazing bifurcations and that these form Arnold tongues in the parameter space spanned by the forcing amplitude and frequency. Further, we show that (i) the wave-less state can regain stability by increasing the driving frequency above the fundamental frequency and (ii) waves can also be generated below this  frequency through nonlinear amplification. Finally, we experimentally observe  hysteresis between a regime of large wave emission and another one where only small waves are radiated;  here too, the boundary of bistability indicates an Arnold tongue pattern.

 In the experiment, one end of the rope, at $x=0$, is initially lifted to a height $Z_0$ and then made to oscillate vertically and harmonically around that position. The elevation $w(x,t)$ of the rope that is not in contact with the ground obeys the beam equation~\cite{Howell-2009}
\beq
\rho \pd{^2w}{t^2}=T \pd{^2w}{x^2}-B \pd{^4w}{x^4}-\rho g,
\label{eq:beam}
\eeq
where $\rho$ is the line density, $B$ is the bending stiffness and $g$ is the acceleration due to gravity. We assume that the rope is under tension,  $T$. This tension could naturally arise from static friction with the ground, as one lifts one extremity of the rope, or it could directly be  applied, as in our experimental set-up. Before shaking the rope, we manually apply a gentle tension on the rope, on the order of $1$\,N, and maintain the resulting static deformation by placing a weight at the unlifted end of the rope (see Fig.~\ref{fig:schematics}). This is done in order to maintain a small slope, which is required for the validity of Eq.~(\ref{eq:beam}). In practice, due to space constraint, the rope made an angle of approximately 30$^\circ$ with the horizontal at the actuated end. In what follows, we assume for simplicity that $T$ is kept constant and uniform in $x$, which neglects variations associate to stretch or slip along the ground and requires the slope to remain small~\cite{Coleman-1992}. This is in contrast with recent studies on the formation of rucks~\cite{Kolinski-2009,Vella-2009} where that force is deduced through the constrain of inextensibility. Note that rucks are essentially under compression rather than in tension, leading to a completely different dynamics.   At $x=0$, one has 
\begin{align}
w&=Z_0 +\eps \Delta Z(t) , &\Delta Z(t)/Z_0&=\cos \omega t, &\eps\ll1,
\label{eq:forcing}
\end{align}
and, in the absence of applied moment, $\pd{^2w}{x^2}=0$. At the contact point $x=x_c(t)$, the boundary conditions are $w=\pd wx=\pd{^2w}{x^2}=0$.   Our aim is to describe the rope dynamics as a function of $\eps$ and $\omega$.

In spite of appearances, the above differential problem is strongly nonlinear: The solution $w(x,t)$ of Eq.~(\ref{eq:beam}) nonlinearly depends on $x_c(t)$, which, in turn is a functional of $w(x,t)$. A hint of nonlinear behaviour can already be found from travelling-wave solutions of Eq.~(\ref{eq:beam}) far away down the rope in that their speed depends on their amplitude~\cite{Howell-2009}. Before embarking on the analysis, we note that a considerable simplification of Eq.~(\ref{eq:beam}) can be made. In our experiments, we use stranded wires. We estimate
their bending rigidity by measuring the torque required to make them conform to a quarter of circle of prescribed radius. For the thickest wire (diameter 6mm), we find $B\approx 0.01$\,Nm$^2$. Hence, over a length scale of 1\,m and with $T\approx1$\,N, one clearly has $B \pd{^4w}{x^4}\ll T \pd{^2w}{x^2}$ and the bending term can safely be neglected everywhere except in boundary layers near $x=0$ and $x=x_c$. Therefore, even though we numerically simulate Eq.~(\ref{eq:beam}) we focus, for the sake of analytical investigation, on
\begin{align}
\pd{^2w}{t^2}&=c^2 \pd{^2w}{x^2} -  g, &c^2=T/\rho,
\label{eq:rope}
\end{align}
 subjected to Eq.~(\ref{eq:forcing}) and only $w=\pd wx=0$ at $x=x_c$. Expanding $x_c$ as
\beq
x_c\sim x_0+\eps x_1+\eps^2 x_2+\ldots ,
\label{expand}
\eeq
the boundary conditions can be expanded as
\beq
Y(x_0,t)+\left(\eps x_1+\eps^2x_2\right)\left.\pd{Y}{x}\right|_{x_0}+\eps^2 \frac{x_1^2}2\left.\pd{^2Y}{x^2}\right|_{x_0}+\ldots=0,
\label{BC3}
\eeq
where $Y=w,\pd{w}{x}$. Furthermore, $w$ is also expanded as
\beq
w\sim w_0+\eps w_1+\eps^2 w_2+\ldots.
\eeq
Substituting the above expansions into Eq.~(\ref{eq:rope}), the leading-order problem is a static one and is easily solved:
\begin{align}
w_0&=\frac{g\left(x-x_0\right)^2}{2c^2}, & x_0^2&=\frac{2c^2Z_0}g=\frac{2TZ_0}{\rho g}.
\label{leading}
\end{align}
At the next order, we have $\left(\pd{^2 }{t^2}-c^2\pd{^2 }{x^2}\right)w_1=0$, which has the general solution $w_1=F(t-x/c)+G(t+x/c)$. Using Eq.~(\ref{eq:forcing}), one has $F(t)+G(t)=\Delta Z(t)$. Next, the two boundary conditions (\ref{BC3}) yields $F(t-x_0/c)+G(t+x_0/c)=0$ and $-F'(t-x_0/c)+G'(t+x_0/c)=-gx_1(t)/c$. Eliminating the functions $F$ and $G$ from these three equations, we find
\beq
x_1(t)-x_1(t-2x_0/c)=(2c/g)\Delta Z'(t-x_0/c).
\label{linear}
\eeq
The above equation describes the linear response of the contact point to a general small-amplitude excitation $\eps\Delta Z(t)$. In the particular case of a harmonic excitation, $\Delta Z/Z_0= \cos\omega t$, one easily finds, using Eq.~(\ref{leading}), that  
\beq
x_1=\frac{x_0 \cos\omega t}{2\sinc(\omega x_0/c)},
\label{linresponse1}
\eeq
 where $\sinc(x)=\sin(x)/x$. An infinite linear response is thus found at frequencies $\omega/(2\pi)=f_p$ where
\begin{align}
 f_p&=p\times\sqrt{\frac{g}{8Z_0}}, &p=1,2,\ldots.
 \label{resonance1}
\end{align}
The set of these resonances make the rope equivalent to a  resonator of length $x_0$, in which waves can travel at speed $c$. What is surprising, however, is that the tension $T$ is absent from the expression of the resonances. This is because both $c$ \emph{and} $x_0$ increase in proportion to $\sqrt T$. The fundamental resonance is akin to that of a classical pendulum of length $2Z_0/\pi^2$. Alternatively, Eq.~(\ref{eq:beam}) can be recast in dimensionless form to show that $T$ disappears from the mathematical formulation in the limit $B/(Tx_0^2)\to0$.  Indeed, with $\xi=x/x_0$, $W=w/Z_0$ and $\tau=ct/x_0$, Eq.~(\ref{eq:beam}) becomes 
\begin{align}
\pd{^2W}{\tau^2}&=\pd{^2W}{\xi^2}-2-\beta \pd{^4W}{\xi^4}, &\beta=\frac{B}{Tx_0^2}=\frac{\rho gB}{2T^2Z_0}
\label{eq:beam2}
\end{align}
with $W-1-\eps\cos\Omega \tau=W_\xi=0$ at $\xi=0$ and $W=W_\xi=W_{\xi\xi}=0$ at $\xi=\xi_c(\tau)$, and 
\beq
\Omega= \omega x_0/c= \omega \sqrt{ 2Z_0 /g}.
\eeq
In the $\beta\to0$ limit, the tension is thus scaled out of the problem. For non zero $\beta$, the frequencies $f_p$ are not equispaced anymore and $f_1$ increases slightly, up to 42\% as $\beta\to\infty$ (see Supplemental Material).

We have checked the independence of $f_1$ on $T$ experimentally. We mounted a stepper motor (Nema 23) capable of delivering a torque of up to 3\,Nm. The motor was driven by a QGL-HQ MA860H pilot, whose signal came from an Arduino Mega2560. The rotation of the motor was converted into vertical motion using a home-made Scott Russel linkage. With this set-up, a precise motion $\Delta Z(t)$ could be imparted on the rope. We used a thin stranded metallic wire rope (diameter 2mm) with $\rho=15$\,g/m, $B\approx 0.0005$\,Nm$^2$. The static elevation $Z_0$ was 0.81\,m, yielding a theoretical fundamental resonance $\sqrt{g/8Z_0}\approx 1.23$\,Hz. Note that the slope of the rope is not very small at $x=0$. Nevertheless, the rope rapidly becomes horizontal as it nears the ground. Additionally, we found that the static profile was almost undistinguishable from a quadratic one, as in Eq.~(\ref{leading}), away from the immediate vicinity of the origin. Eqs.~(\ref{eq:beam}) and (\ref{eq:rope}) thus appear to be reliable.  In order to avoid sideways motion of the rope, also known as ``ponytail instability''~\cite{Belmonte-2001,Keller-2010}, the rope was vertically oscillated against a board. This produced some friction, but much less than the internal one. According to the linear theory, a general driving $\Delta Z(t)/Z_0=\int  A(\omega) \exp(i\omega t)\rd \omega$, yields a linear response
\beq
x_1(t)=\frac{x_0}{2}\int \frac{ A(\omega) e^{i\omega t}}{\sinc(\omega x_0/c)}\,\rd \omega.
\label{linresponse2}
\eeq
Hence, in order to excite the fundamental resonance only, $\Delta Z(t)$ was gradually ramped from a vanishing amplitude to a constant sinusoidal excitation in 20 periods of oscillation. In this way, $A(\omega)$ was strongly peaked around a well-defined frequency and unwanted excitation of high order poles in Eq.~(\ref{linresponse2}) was reduced. We increased the frequency until resonance was detected, in the form of a transition to small waves being emitted down the rope. Because of the very small amplitude of the waves, there was some uncertainty on the frequency at the transition on the order of 0.02\,Hz. We repeated the experiment with different values of the tension $T$, measured by the static part $x_0$ of the contact point. To measure $T$ directly was difficult to implement but, from Eq.~(\ref{leading}), $x_0$ increases monotonically with $T$. Independence of $f_1$ on $x_0$ thus implies independence of $T$, all other parameters being unchanged. The results, shown in Table.~\ref{tab:exp:resonance}, confirm the prediction.

\begin{table}[h]\caption{\label{tab:exp:resonance}Experimental demonstration of independence of resonance frequency on the tension in the rope. $\rho=15$\,g/m, $B\approx0.0005$\,Nm$^2$, $\eps=0.05$, $Z_0=81$\,cm, $\sqrt{g/8Z_0}\approx 1.23$\,Hz.}\begin{ruledtabular}
\begin{tabular}{llllllll}
$x_0\pm0.03$ (m): &1.08 &1.20 &1.47& 1.65 &1.80&2.30&2.45\\
$f_1\pm0.02$ (Hz): &1.18& 1.19& 1.20& 1.20& 1.23& 1.17& 1.20\\
\end{tabular}\end{ruledtabular}\end{table}

Eq.~(\ref{linresponse1}) indicates that, starting from a small frequency below $f_1$ and increasing it, the response of the system diverges. One can anticipate that the unbounded growth of the oscillation can only resolve itself into a train of waves along the rope. On the other hand, in the range $f_1<\omega/(2\pi)<f_2$, the linear response becomes small again, suggesting that it is possible not to emit waves there. However, apart from the divergence, the above linear theory fails to give details on how waves are actually generated. Proceeding to higher orders of the analysis, we  use the coordinate $s=(x-x_c(t))/x_0$ and the reduced pulsation $\Omega=\omega x_0/c$. For a purely harmonic forcing, we find  (See Supplemental Material) that
\beq
w/Z_0\sim\sum_{j\geq0}\left(\frac{\eps}{\sin\Omega}\right)^jF_j(s,\eps) \cos\left(j\omega t\right),
\label{expandw}
\eeq
where $F_0\sim s^2+O(\eps^2)$,
\begin{multline}
F_1\sim \left(\Omega s-\sin\Omega s\right)\left[1+\frac{\eps^2\Omega^4\cos^2\Omega}{32\sin^4\Omega}\left(1+2\frac{\tan\Omega}{\Omega}\right)\right]
\\
+\frac{\eps^2\Omega^4\sec\Omega}{16\sin^3\Omega} 
\left[\sin2\Omega-\sin\left(2\Omega+\Omega s\right)\right]\sin\Omega s 
+O\left(\eps^4\right)
\end{multline}
\begin{multline}
F_2\sim\frac{\Omega^2}{8}\Big[\left(1-\cos\Omega s\right)^2-\sin^2\Omega s-\\
\frac{\cos2\Omega}{\sin2\Omega}\left(2\Omega s-\sin2\Omega s\right)\Big]+O\left(\eps^2\right)
\end{multline}
and
\begin{multline}
F_3\sim\frac{\Omega^4}{32}
\Big[\sin\Omega s+\sin3\Omega s-2\sin2\Omega s
\\
+
\frac{\left(2-\cos2\Omega+\cos4\Omega\right)}{2\left(1+2\cos2\Omega\right)\sin^2\Omega}\left(3\Omega s-\sin3\Omega s\right)
 \\
-4\frac{\cos2\Omega}{\sin2\Omega}\left(1-2\cos\Omega s\right)\sin^2\Omega s
\Big]+O\left(\eps^2\right)
\label{F3}
\end{multline}
The leading order expressions of $F_2$ and $F_3$ indicate new resonances when $\sin2\Omega=0$ and when  $1+2\cos2\Omega=0$, that is when $\omega/(2\pi)=f_p/2$ or $f_p/3$. Indeed, the nonlinearity  in Eq.~(\ref{BC3})  leads to higher harmonics of the  $\cos\omega t$ forcing which, in turn, can match the fundamental  resonances. As one progresses to higher orders in the analysis, more harmonics are found. In particular, the condition $q\omega/(2\pi)=f_p$ yields the nonlinear resonances $\omega_{p,q}/(2\pi)=f_p/q=(p/q)f_1$, similarly to~\cite{Mokhtar-2017}. Since $\{\omega_{p,q}| p,q\in\mathbb Z\}$ is a dense set, any frequency is arbitrarily close to a nonlinear resonance and any harmonic forcing should in principle lead to a divergent response in the absence of dissipation.

\begin{figure}
\includegraphics[width=8cm]{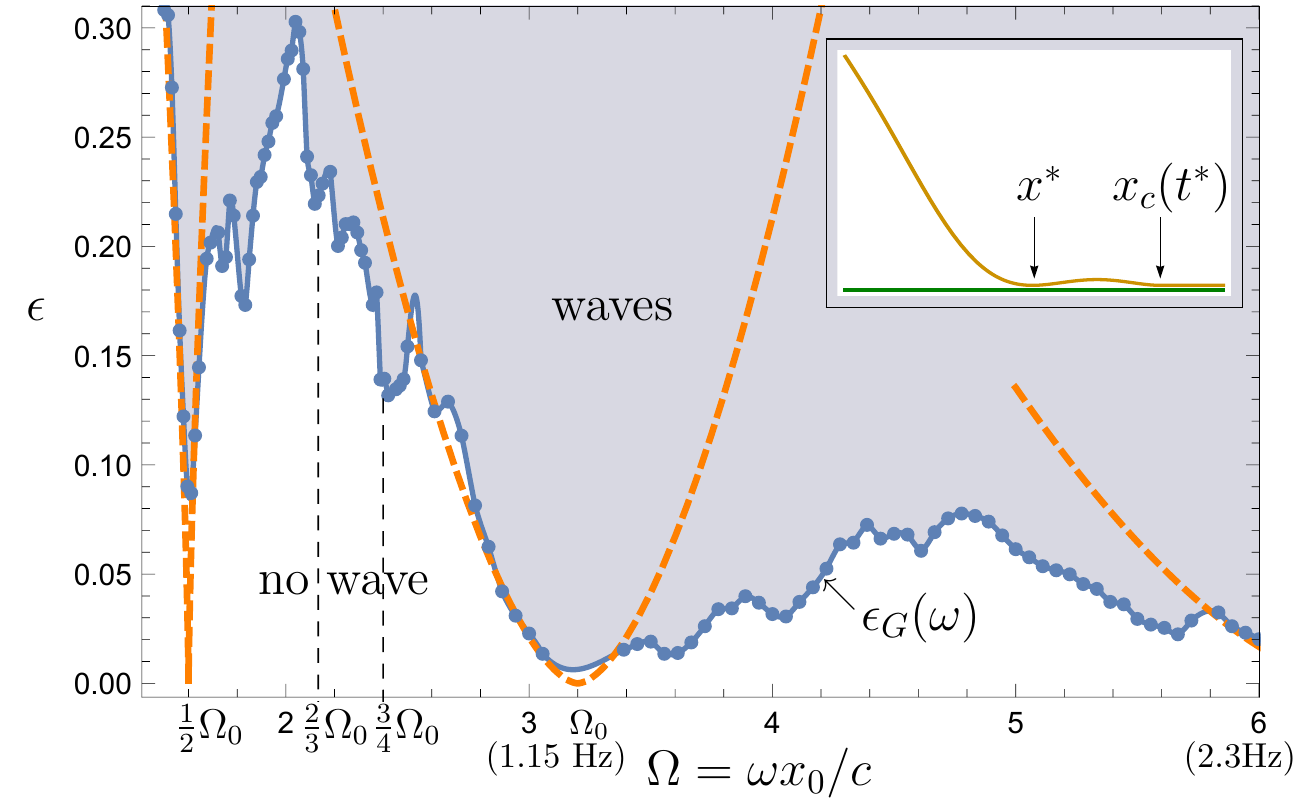}
\caption{Bifurcation diagram in the $(\omega,\eps)$ plane obtained by simulation of Eq.~(\ref{eq:beam}) with $B/(Tx_0^2)=0.01$. Orange dashed curves, from left to right: $\eps= 1.815|\Omega-\Omega_0/2|^{3/4}$, $\eps=0.309|\Omega-\Omega_0|^{5/3}$, and  $\eps=0.077|\Omega-2\Omega_0|^{5/3}$, where $\Omega_0=3.2$. Inset: analytical profile Eq.~(\ref{expandw2}) at impact near the $\Omega_0$ resonance. Between parentheses are indicated frequencies in Hz corresponding to an experiment with $Z_0=81$\,cm.}
\label{fig:compare}
\end{figure}

Let us study the limit $\omega/(2\pi)\to f_p$ of the above expansion. Writing $\Omega=p\pi+\mu$, $\mu\ll1$, Eqs.~(\ref{expandw})-(\ref{F3}) yield, for sufficiently small $\mu$:
\begin{multline}
w/Z_0\sim s^2+(-)^p\frac{\eps^3\left(p\pi\right)^4}{32\mu^5}\bigg[ 
\cos\left(\omega t\right)\left(p\pi s-\sin p\pi s\right)\\ \left.
+\frac13\cos\left(3\omega t\right)\left(3p\pi s-\sin3p\pi s\right)\right].
\label{expandw2}
\end{multline}
This function displays a local minimum at times $t$ given by $\omega t=(2n+p)\pi$. One finds that, at such times, the rope makes contact with the ground at $s=-1.13/(p\pi)$ with zero slope and zero velocity if
\beq
\eps=\eps_G(\omega)\approx\frac{0.309}{p^2} \mu^{5/3}=\frac{0.309 }{p^2} \left(\frac{\omega x_0}{c}-p\pi\right)^{5/3}.
\label{grazing}
\eeq
Importantly, Arnold tongues given by the above formula become flatter as $p$ increases. Hence, the areas below these curves shrink as $1/p^2$. 
 Similarly, using the $O\left(\eps^4\right)$ expression for $F_4(s,\eps)$ (see Supplemtal Material), one may investigate the region $\omega/(2\pi)\to f_1/2$, by writing $\Omega=\pi/2+\mu$, $\mu\ll1$. One then obtains that the rope touches the ground before $x=x_0$ if, locally,
\beq
\eps=\eps_G(\omega)\approx1.815\mu^{3/4}=1.815\left(\frac{\omega x_0}{c}-\frac\pi2\right)^{3/4}. 
\eeq

\begin{figure*}
\includegraphics[width=.8\textwidth]{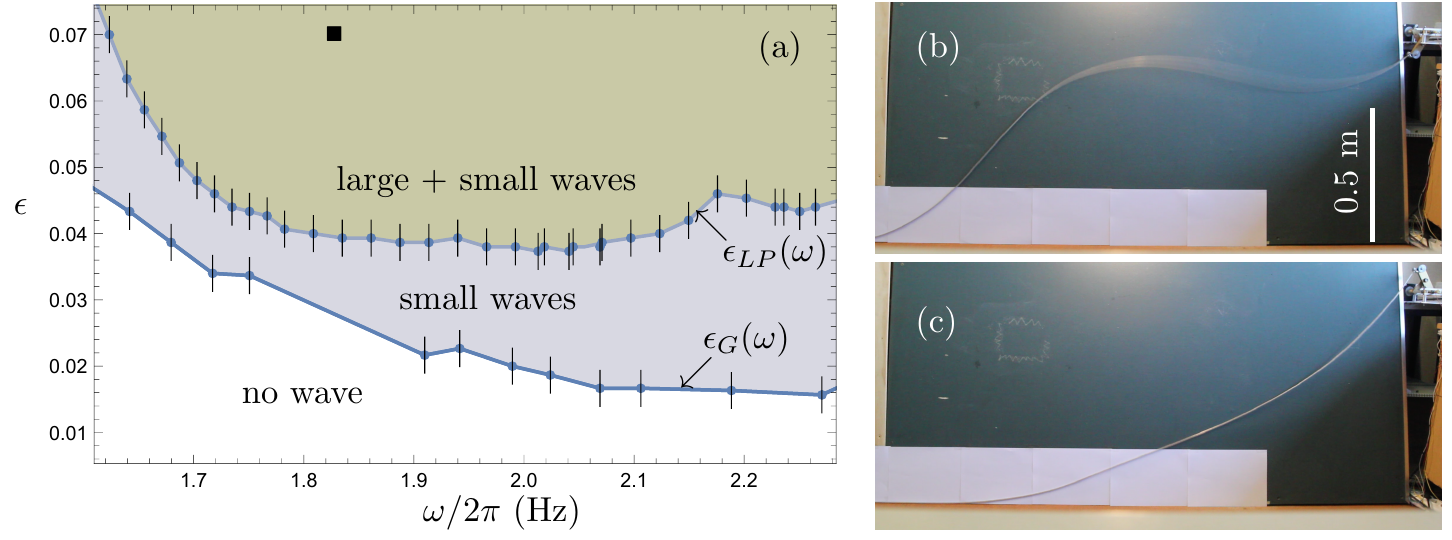}
\caption{(a) : Experimental bifurcation diagram in the $(\omega,\eps)$ parameter plane. $\eps_{LP}(\omega)$: limit-point curve above which large-wave and small-wave regimes coexist. $\eps_{G}(\omega)$: grazing bifurcation curve. (b) : snapshots of the large-wave oscillation. (c): snapshot of the small-wave oscillation (extracted frames from video in Supplemental Material.) Panels (b) and (c) both correspond to the same value of $\omega$ and $\eps$ but different initial conditions [black square in panel (a).] $\rho=48$\,g/m, $B\approx0.01$\,Nm$^2$, $Z_0=81$\,cm}
\label{fig:experiment}
\end{figure*}

Eq.~(\ref{expandw2}) is not rigorously valid because it breaks the asymptotic ordering of terms in Eq.~(\ref{expandw}). Therefore, it shouldn't be given any theoretical value other than providing a qualitative trend. Nevertheless, the resulting expression Eq.~(\ref{grazing}) is found to fit remarkably well with the appearance of waves in numerical simulations. Numerical simulations of dynamical contact problems are known to be challenging~\cite{Doyen-2011}. A practical issue is the determination and proper processing of the contact set in the discretized rope. We used an algorithm adapted from Liakou~\textit{et al.}~\cite{Liakou-2017}. A description is given in the Supplemental Material.
 Fig.~\ref{fig:compare} shows the boundary between the regimes of no radiation and radiation.  Numerically, the fundamental resonance is slightly shifted from $\Omega=\pi$ to $\Omega=\Omega_0\approx3.2$; this results from the small but non zero bending stiffness [$B/(Tx_0^2)=0.01$]. Strikingly, the numerical curve contains a large number of sharp dips, in addition to those predicted by the linear analysis. These are signatures of the nonlinear resonances discussed above. In particular, a sharp resonance is seen at $\Omega_0/2$. As anticipated by the theory,  the wave-less state regains stability in the range $\Omega_0<\omega x_0/c<2\Omega_0$ for sufficiently small $\eps$. The analytical formulas derived above in the vicinity of $\Omega_0/2$, $\Omega_0$ and $2\Omega_0$ convincingly  fit the numerical curve.  


The situation described by Eqs.~(\ref{expandw2}) and (\ref{grazing}) is reminiscent of  a \emph{grazing bifurcation}. Such a bifurcation classically applies to a point mass attached to a spring and subjected to a periodic force. Upon increasing the forcing amplitude, the mass starts making contact with an obstacle. Beyond the grazing bifurcation, the dynamics rapidly becomes chaotic~\cite{Shaw-1983}. The crucial difference here, of course, is that the rope is spatially distributed. Nevertheless, the analogy is sufficiently strong to also call the threshold identified by Eq.~(\ref{grazing}) a grazing bifurcation. Let us denote by $t^*$ and $x^*$ the time and location of impact as determined above. If $\eps$ exceeds the grazing bifurcation threshold, the contact happens with a finite speed, $-V$.  In the limit of an infinitely rigid ground with restitution coefficient $r$, the rope locally rebounds instantaneously with speed $rV$. This amounts to a reaction force $\left(1+r\right)\rho V\delta(t-t^*)\delta(x-x^*)$, which produces an elevation $w^*(x,t)=0.5\left(1+r\right)Vx_0/c$ in the range $-c(t-t^*)<x-x^*<c(t-t^*)$~\cite{Howell-2009}. The perturbation $w^*$ expands in both directions and adds itself to the elevation given by Eq.~(\ref{expandw}). Ultimately, this provokes the detachment of the  the bump between $x^*$ and $x_c(t^*)$ and its propagation at speed $c$ down the rope. In this scenario, the portion of the rope ahead of $x^*$  which ultimately forms the travelling wave, is initially given by Eq.~(\ref{expandw2}) at $t^*$.

We now turn to the experimental demonstration of wave generation through harmonic forcing at $x=0$. As previously mentioned, the computer-driven motor was mounted on a table at a height $Z_0=81$cm. The angle of the rotor varied in steps of 1.8$^\circ$. Given the length of the arm of the Russel linkage, this translates into an uncertainty of $\Delta Z(t)$ of 4mm, hence an uncertainty $|\Delta\eps|\approx0.005$. The transitions  were monitor by varying $\eps$ for fixed $\omega$. Being computer-controlled, the uncertainty in frequency is estimated to be well under 0.01\,Hz and hence, negligible.  From what precedes, waves emitted at the grazing bifurcation point $\eps_G(\omega)$ are of very small amplitude, the maximum being approximately $0.085 Z_0/(p\pi)^2$ (see inset of Fig.~\ref{fig:compare}). Moreover, past the bifurcation threshold, this amplitude does not grow rapidly as $\left(\eps-\eps_G(\omega)\right)^{1/2}$, but, rather, only linearly in $\eps-\eps_G(\omega)$, as in other impact systems~\cite{Bernardo-2008}. Finally, the waves undergo rapid attenuation due to internal friction between the strands of the rope. 
 This makes the determination of the transition experimentally challenging, and we focused on large frequencies, near $2\Omega_0$, in order to benefit from a large amplification of the oscillations. 

Fig.~\ref{fig:experiment}(a) shows the experimental result. The no wave/small wave boundary is the grazing bifurcation curve. It shows a minimum near 2.3\,Hz, in good agreement with the $f_2$ resonance of Fig.~\ref{fig:compare}. However, quantitative agreement between the numerical curve of Fig.~\ref{fig:compare} and the experimental one in Fig.~\ref{fig:experiment}(a) is poor, even though the order of magnitude is the same. We attribute this poor matching to the non-constancy of $T$. Indeed, $T$ is not actively controlled in the experiment. The fact that the slope of the rope is not everywhere small (see Supplemental Material) together with small parasitic slippage on the ground both make $T$ vary in time and space in practice.

Next to the small-wave regime described thus far, we experimentally discover a large-wave regime. The two regimes stably coexist over wide ranges of parameters. Figs.~\ref{fig:experiment}(b) and (c) are snapshot of the two distinct dynamical states, observed for the same values of $\omega$ and $\eps$ [black dot in Fig.~\ref{fig:experiment}(a).] The limit of coexistence between the two regimes is classically given by a curve of limit point $\eps_{LP}(\omega)$ in the $(\omega,\eps)$ space. 

Fortunately the curve  $\eps_{LP}(\omega)$ is much more convenient to determine than $\eps_G(\omega)$ as the transition between the two regimes is visually clear-cut. We record $\eps_{LP}(\omega)$ in the following way: (i) start with a value $\eps>\eps_{LP}(\omega)$ in the state of large-wave emission. (ii) Decrease $\eps$ by small step and wait for the system to  relax to a stable operation. (iii) As soon as $\eps<\eps_{LP}$, the large-wave state irreversably disappears after only a few oscillations, giving way to the small-wave state. The most important feature of the $\eps_{LP}(\omega)$ curve is its minimum. This minimum attests of the resonant nature of that state and is an indirect manifestation of the Arnold tongues described above. 

Regarding the large-wave emitting state, we make two observations. Firstly, we could not observe it with ropes of smaller cross sections, \textit{i.e.} smaller bending stiffness $B$. This suggests that bending stiffness plays an important role in the existence of this dynamical regime. Secondly, we could not reproduce the bistability between a large-wave and a small-wave making state for any value of $B$ in Eq.~(\ref{eq:beam}). From this, we conclude that this model is insufficient to describe this state and that a fully geometrically nonlinear model is required  \cite{Howell-2009,Belmonte-2001,Coleman-1992}.  Geometrical nonlinearity couples the transverse motion $w(x,t)$ to a longitudinal motion, making the displacement fully vectorial. This could be the subject of further investigation.

Dynamical contacts between deformable bodies may display rich dynamical behaviours. Using a simple rope as a prototype example, we have seen that the dynamics combines the physics of free moving boundaries and that of non-smooth dynamical systems. In particular, contacts are governed  by the infinite set of resonant deformation modes of the rope [see the multiple poles in Eq.~(\ref{linear})], supplemented by their nonlinear harmonics. When analyzing more complicated  dynamical contact problems, where analytical results are out of reach, the present study suggests to pay attention to linear resonances and their nonlinear harmonics. In the vicinity of these resonances, nonlinear waves and chaos are susceptible to arise via grazing bifurcations. Finally, the stability regions are likely to be delimited by Arnold tongues in the parameter space. Thanks to the slenderness of the rope, the elastic degrees of freedom are reduced to their simplest expression, here. Next in complexity would be the study of time-dependent contacts between two slender bodies and the inclusion of out-of-plane degrees of freedom.

\acknowledgments
B.S. is a Research Fellow and G.K. is a Research Associate of the Fonds de la Recherche Scientifique - FNRS (Belgium.) We thank Pascal Damman and Fabian Brau  for stimulating discussions and Jamal Tahmaoui for fabricating the Scott Russel linkage.


%

\end{document}